\documentclass[11pt]{article}

\usepackage{graphicx}
\usepackage{dcolumn}
\usepackage{bm}

\begin{document}
\centerline{\bf\Large Study on Form Factors at Effective Vertices}
\vspace{0.3cm}

\centerline{\bf\Large of Diqarks Coupling to Gauge Bosons}

\vspace{1cm}

Yan-Ming Yu$^1$, Hong-Wei Ke$^1$, Yi-Bing Ding$^2$, Xin-Heng
Guo$^3$, Hong-Ying Jin$^4$, Xue-Qian Li$^1$, Peng-Nian Shen$^5$
and Guo-Li Wang$^6$

\vspace{0.5cm}

1. Department of Physics, Nankai University, Tianjin 300071,
China.

2. Department of Physics, The Graduate School of Chinese Academy
of Sciences, Beijing 100049, China.

3. Institute of Low Energy Nuclear Physics, Beijing Normal
University, Beijing 100875, China.

4. Institute of Modern Physics, Zhejiang University, Hangzhou
310027, China.

5. Institute of High Energy Physics, Chinese Academy of Sciences,
P.O. Box 918-4, Beijing 100049, China.

6. Department of Physics, The Harbin Polytechnical University,
Harbin 150001,  China.

\vspace{1cm}

\begin{center}
\begin{minipage}{12cm}
\noindent{Abstract:}

The diquark structure in baryons is commonly accepted as a
reasonable approximation which can much simplify the picture of
baryons and reduce the length of calculations. However, a diquark
by no means is a point-like particle, even though it is treated as
a whole object. Therefore, to apply the diquark picture to
phenomenological calculations, at the effective vertices for the
diquark-gauge boson interactions, suitable form factors must be
introduced to compensate the effects caused by the inner structure
of the diquark. It is crucial to derive the appropriate form
factors for various interactions. In this work, we use the
Bethe-Salpeter equation to derive such form factors and
numerically evaluate their magnitudes.

\end{minipage}
\end{center}

\section{Introduction}

In the quark model, regular baryons are composed of three valence
quarks. Compared with the case for meson which contains a quark
and an antiquark, the physical picture for baryon is much more
difficult to deal with because a three-body system is terribly
more complicated than a two-body system. It is believed that the
correct description for a baryon which contains three quarks is
the Faddeev equation\cite{faddeev} whereas the Bethe-Salpeter
equation properly describes the meson structure.

The concept of diquark was raised even at the epoch of the birth
of the quark model \cite{gellman}. However, it is still in dispute
that diquark is a substantial structure of color-anti-triplet or
just a mathematical decomposition of the  $3\otimes 3\otimes 3$
representation of $SU_c(3)$ into $3\otimes \bar 3$. For the
baryons which are composed of two heavy quarks (b,c) and a light
quark, it is believed that the two heavy quarks can constitute a
more stable diquark with smaller size, but for the baryons with
one-heavy-two-light-quark structure or even three-light-quark
structure, the diquark-picture is dubious. Analyzing the Faddeev
equation, one finds that the diquark picture is an approximation
where two quarks are supposed to constitute a more stable
subsystem and the interaction with the other quark can be treated
as a perturbation which is not strong enough to break the diquark
binding.

For a long time, the concept of diquarks has been applied to study
the processes where baryons are involved
\cite{ida,lichtenberg,bose1,bose2,miyazawa,anselmino,jandw,jaffe,wilczek,shuryak,zou,wang,roberts}.
With the diquark-quark sructure, the physical picture of baryons
is simplified and the theoretical calculations become much easier.
However,  one would ask what is missing in the simplified version.
In other words, the diquark is by no means a real point-like
particle, but has inner structure. For lower energy, the transfer
momentum is small, then the inner structure may not manifest
itself and the diquark can be nicely treated as a point. On the
contrary, as the energy scale in the problem is larger, simply
treating it as a structureless particle would bring up large
errors to the theoretical calculations. Thus one needs to involve
such effects which reflect the inner structure of diquarks. One
introduces form factors at each effective vertices where diquarks
interact with gauge bosons, such as $W^{\pm},\; Z,\; \gamma$ and
gluons, to take into account the effects caused by the inner
structure. Since diquark is generally in color anti-triplet, and
not a physical object, the form factors cannot be experimentally
measured, thus one needs to use sort of theoretical models to
derive them. The phenomenological expressions of such form factors
were introduced by many authors
\cite{anselmino1,guo,kroll,keiner}. The authors of
Refs.\cite{maris,gerasyuta} used the Dyson-Schwinger equation to
evaluate the electromagnetic form factors of diquarks. Guo et al.
\cite{guo1} used the heavy quark effective theory to derive the
form factor for the diquark-gluon coupling and Ebert et
al.\cite{ebert} employed the relativistic quark model to obtain
these form factors. Dai\cite{dai} et al. calculated baryon
transitions.  Ahlig et al. \cite{ahlig} derived the wavefunction
for the baryon-diquark-quark couplings.

Indeed, as the diquark picture is accepted, it is necessary to
derive the form factors at the effective vertices for
diquark-gauge-boson couplings or even diquark-mesons couplings,
because diquarks are not point-like particles and the effects of
the inner structure of diquarks must manifest themselves through
the form factors. In this work, we try to derive the form factors
at the effective vertices for diquark-gauge-boson couplings in a
more general framework and the form factors may be applied to
phenomenological calculations of decay rates or production
processes where baryons, especially heavy baryons are involved.

The framework we are going to adopt is the Bethe-Salpeter equation
(BSE). As discussed above the BSE may be a simplified version of the
complicated Faddeev equation which seems to well describe the baryon
structure. The BSE established on the quantum field theory is
considered to be a feasible approach to study the relativistic
two-body bound states. Salpeter \cite{salpeter} adopted the
instantaneous approximation to simplify the BSE which can be applied
to deal with practical phenomenological problems. Namely, to
determine the form factors, we can first obtain the diquark spectra
and wavefunctions \cite{yu} and then with them as inputs we adopt
the instantaneous BSE method developed by Chang et al. \cite{chang1}
to derive the form factors at the effective vertices of diquarks
coupling to gauge bosons: gluon, photon, $W^{\pm}$ and $Z^0$. Then
we also try to extend the method to obtain the effective vertex of
diquark coupling to pseudoscalar mesons, such as pions.

In this work, after the introduction, we derive all the form
factors at the effective vertices in terms of the BSE in Sec.II.
In Sec.III the numerical results are presented while the input
parameters are given explicitly, and then the last section is
devoted to the summary and discussions.

\section{Formulation}
\subsection{The form factors of diquark coupling to gluons}
The effective vertex of diquark coupling to a gauge boson
($g,\;\gamma,\; Z^0$) can be written in a form as
\begin{equation}
V^{eff}_{DGD}=\Gamma^{\mu}_{DGD}(V_{qGq})\cdot\epsilon_{\mu},
\end{equation}
where $\epsilon_{\mu}$ is the polarization vector of the gauge
boson, $D,\;G$ refer to the diquark and the gauge boson,
respectively. Obviously, $\Gamma^{\mu}_{DGD}$ is a unique function
of the quark-gauge-boson vertex $V_{qGq}$ which is given in the
fundamental theories. Here we only concern the standard model
(SM). Below, we are going to derive all the currents
$\Gamma_{DGD}^{\mu}$ in terms of the BSE.

1. The effective current of scalar diquark coupling to gluons can
be written as:
\begin{eqnarray}\Gamma^{\mu,a}_{sgs}=-ig_s\frac{\lambda^a}{2}G(Q^2)(P_f+P_i)^\mu,
\end{eqnarray}
\begin{eqnarray}G(Q^2)=\frac{P_{x\mu}[M^\mu_1+M^\mu_2]}{(P_f+P_i)\cdot
P_x},
\end{eqnarray}
where $G(Q^2)$ is the form factor, $P_i$ and $P_f$ are the momenta
of the initial and final diquarks respectively and
$Q^2=(P_f-P_i)^2$ is the momentum transfer, $P_x$ is an arbitrary
non-zero auxiliary four-vector. The physical picture of a diquark
coupling to a gauge boson is depicted in Fig. 1, and the sum of
Fig.1 (a) and (b) makes the net contribution to the form factor.

\begin{figure}[htb]
\begin{center}
\mbox{ \scalebox{0.8}{\includegraphics{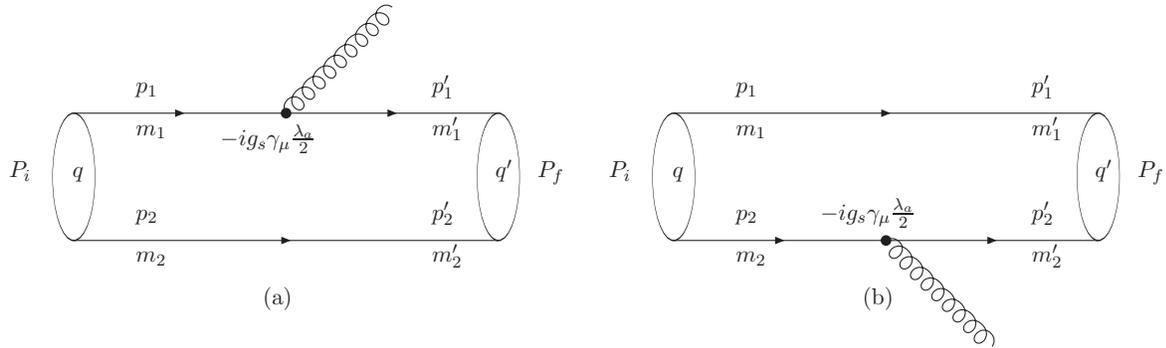}}
 }
\end{center}
    \caption{The Feynman diagram for diquark coupling to a gauge boson}
    \label{pic1}
\end{figure}

The effective current corresponding to Fig. 1 (a) is
\begin{eqnarray}M^\mu_1&&=-\int\frac{d^4qd^4q'}{(2\pi)^4}\Big\{\delta^4(p_2-p'_2)Tr\Big[\overline{\chi}_{_{P_f}}(q')
\gamma^\mu\chi_{_{P_i}}(q)
S^{-1}_F(-p_2)\Big]\Big\}\nonumber\\
&&=-\int\frac{d^4q}{(2\pi)^4}Tr\Big[\overline{\chi}_{_{P_f}}(q')
\gamma^\mu\chi_{_{P_i}}(q) S^{-1}_F(p_2)\Big]\label{eq1}.
\end{eqnarray}
Here $p_2=\alpha_2P_i-q$, $p'_2=\alpha_2P_f-q'$, and by
conservation of momentum one has
\begin{eqnarray}
q'=\alpha_2(P_f-P_i)+q,
\end{eqnarray}
and the current corresponding to Fig.1 (b)  is
\begin{eqnarray}
M^\mu_2&&=-\int\frac{d^4qd^4q'}{(2\pi)^4}\Big\{\delta^4(p_1-p'_1)Tr\Big[\overline{\chi}_{_{P_f}}(q')S^{-1}_F(p_1)
\chi_{_{P_i}}(q)\gamma^\mu\Big]\Big\}\nonumber\\
&&=-\int\frac{d^4q}{(2\pi)^4}Tr\Big[\overline{\chi}_{_{P_f}}(q')S^{-1}_F(p_1)
\chi_{_{P_i}}(q)\gamma^\mu\Big],\label{eq2}
\end{eqnarray}
where $p_1=\alpha_1P_i+q$, $p'_1=\alpha_1P_f+q'$ and
\begin{eqnarray}
q'=\alpha_2(P_i-P_f)+q.
\end{eqnarray}
By the BSE and under the instantaneous approximation, we obtain,
\begin{eqnarray}\label{gs1}\chi_{_{P_i}}(q)&&=\frac{1}{p\!\!\!\slash_1-m_1}\int
d^3k_{_{P_T}}\Big[V(|q_{_{P_\perp}}-k_{_{P_\perp}}|)\varphi_{_{P_i}}(k_{_{P_\perp}})\Big]
\frac{1}{p\!\!\!\slash_2+m_2}\nonumber\\
&&=\frac{1}{p\!\!\!\slash_1-m_1}\eta_{_{P_i}}(q_{_{P_\perp}})\frac{1}{p\!\!\!\slash_2+m_2},
\end{eqnarray}
and
\begin{eqnarray}\label{9}
\overline{\chi}_{_{P_f}}(q')&&=\frac{1}{p\!\!\!\slash'_2+m_2}\int
d^3k_{_{P_T}}\Big[\overline{\varphi}_{_{P_f}}(k_{_{P_\perp}})V(|q'_{_{P_\perp}}
-k_{_{P_\perp}}|)\Big]\frac{1}{p\!\!\!\slash'_1-m_1}\nonumber\\
&&=\frac{1}{p\!\!\!\slash'_2+m_2}\overline{\eta}_{_{P_f}}(q'_{_{P_\perp}})\frac{1}{p\!\!\!\slash'_1-m_1},
\end{eqnarray}
where $P$ is the momentum of the baryon which contains the
diquark. In fact, as one only discusses the diquark system, the
baryon momentum is irrelevant. Here we set the baryon momentum $P$
as a reference momentum and then we can properly specify other
momenta, $q^\mu_{_{P_\parallel}}=\frac{q\cdot P}{M^2}P^\mu$,
$q^\mu_{_{P_\perp}}=q^\mu-q^\mu_{_{P_\parallel}}$,
$q_{_P}=\frac{q\cdot P}{M}$ and
$q_{_{P_T}}=\sqrt{q^2_{_P}-q^2}=\sqrt{-q^2_{_{P_\perp}}}$ are the
projections of the inner momentum $q$ of quarks inside the diquark
on the directions parallel and perpendicular to $P$ and
corresponding invariants respectively. $\varphi_{_{P_i(P_f)}}$ in
Eqs. (\ref{gs1},\ref{9}) are defined as
\begin{eqnarray}\varphi_{_{P_i(P_f)}}(q_{_{P_\perp}})=\int
dq_{_P}\chi_{_{P_i(P_f)}}(q_{_P},q_{_{P_\perp}}).
\end{eqnarray}
The definitions of the subscripts are obvious. $S^{-1}_F(p_1)$ is
the inverse of a fermion propagator
$\frac{1}{p\!\!\!\slash'_1-m_1}$, and from Eq.(\ref{gs1}), one can
note that on the other leg connecting to the kernel, there should
be another fermion propagator corresponding to $p_2$. However, as
we properly convert the BS wavefunction into a $4\times 4$ matrix
form from a $16\times 1$ matrix \cite{greiner}, it automatically
turns into its charge conjugation which is equivalently expressed
as $\frac{1}{p\!\!\!\slash'_2+m_2}$. Then we can re-write Eqs. (4)
and (6) as
\begin{eqnarray}M^\mu_1=-i\int\frac{d^4q}{(2\pi)^4}Tr\Big[\frac{1}{p\!\!\!\slash'_2+m_2}
\overline{\eta}_{P_f}(q'_{_{P_\perp}})\frac{1}{p\!\!\!\slash'_1-m_1}\gamma^\mu\frac{1}{p\!\!\!\slash_1-m_1}
\eta_{P_i}(q_{_{P_\perp}})\Big],
\end{eqnarray}
and
\begin{eqnarray}M^\mu_2=-i\int\frac{d^4q}{(2\pi)^4}Tr\Big[\frac{1}{p\!\!\!\slash'_2+m_2}
\overline{\eta}_{P_f}(q'_{_{P_\perp}})\frac{1}{p\!\!\!\slash'_1-m_1}\eta_{P_i}(q_{_{P_\perp}})
\frac{1}{p\!\!\!\slash_2+m_2}\gamma^\mu\Big].
\end{eqnarray}
Following the commonly adopted method, we also decompose the
propagators as
\begin{eqnarray}\frac{1}{p\!\!\!\slash_1-m_1}=\frac{\Lambda^+_1}{\alpha_1P_{iP}+q_{_{P}}-\omega_1
+i\epsilon}+\frac{\Lambda^-_1}{\alpha_1P_{iP}+q_{_{P}}+\omega_1-i\epsilon},
\end{eqnarray}
\begin{eqnarray}\frac{1}{p\!\!\!\slash_2+m_2}=\frac{\Lambda^+_2}{\alpha_2P_{iP}-q_{_{P}}-\omega_2
+i\epsilon}+\frac{\Lambda^-_2}{\alpha_2P_{iP}-q_{_{P}}+\omega_2-i\epsilon},
\end{eqnarray}
\begin{eqnarray}\frac{1}{p\!\!\!\slash'_1-m_1}=\frac{\Lambda'^+_1}{\alpha_1P_{fP}+q'_{_{P}}
-\omega'_1+i\epsilon}+\frac{\Lambda'^-_1}{\alpha_1P_{fP}+q'_{_{P}}+\omega'_1-i\epsilon},
\end{eqnarray}
\begin{eqnarray}\frac{1}{p\!\!\!\slash'_2+m_2}=\frac{\Lambda'^+_2}{\alpha_2P_{fP}-q'_{_{P}}
-\omega'_2+i\epsilon}+\frac{\Lambda'^-_2}{\alpha_2P_{fP}-q'_{_{P}}+\omega'_2-i\epsilon},
\end{eqnarray}
where $\alpha_i\equiv \frac{m_i}{m_1+m_2}\;(i=1,2)$ and
\begin{eqnarray}&&\omega_1=\sqrt{m^2_1-(\alpha_1P_{iP_\perp}+q_{_{P_\perp}})^2},\\
&&\omega_2=\sqrt{m^2_2-(\alpha_2P_{iP_\perp}-q_{_{P_\perp}})^2},
\end{eqnarray}
\begin{eqnarray}\Lambda^\pm_1=\frac{\frac{P\!\!\!\slash}{M}\omega_1\pm(m_1+q\!\!\!\slash_{_{P_\perp}}
+\alpha_1P\!\!\!\slash_{iP_\perp})}{2\omega_1},
\end{eqnarray}
\begin{eqnarray}\Lambda^\pm_2=\frac{\frac{P\!\!\!\slash}{M}\omega_2\mp(m_2+q\!\!\!\slash_{_{P_\perp}}
-\alpha_2P\!\!\!\slash_{iP_\perp})}{2\omega_2},
\end{eqnarray}
\begin{eqnarray}&&\omega'_1=\sqrt{m^2_1-(\alpha_1P_{fP_\perp}+q'_{_{P_\perp}})^2},\\
&&\omega'_2=\sqrt{m^2_2-(\alpha_2P_{fP_\perp}-q'_{_{P_\perp}})^2},
\end{eqnarray}
\begin{eqnarray}\Lambda'^\pm_1=\frac{\frac{P\!\!\!\slash}{M}\omega'_1\pm(m_1+q\!\!\!\slash'_{_{P_\perp}}
+\alpha_1P\!\!\!\slash_{fP_\perp})}{2\omega'_1},
\end{eqnarray}
\begin{eqnarray}\Lambda'^\pm_2=\frac{\frac{P\!\!\!\slash}{M}\omega'_2\mp(m_2+q\!\!\!\slash'_{_{P_\perp}}
-\alpha_2P\!\!\!\slash_{fP_\perp})}{2\omega'_2}.
\end{eqnarray}
Substituting these equations into the expressions of $M_1^{\mu}$
and $M_2^{\mu}$ in Eqs. (11, 12), writing $d^4q$ in the covariant
form as $dq_{_P}d^3q_{_{P_T}}$, and completing the contour
integration with respect to $dq_{_P}$, we obtain the following
equation,
\begin{eqnarray}
&&M^\mu_1=\int\frac{d^3q_{_{P_\perp}}}{(2\pi)^3}Tr\Big\{-\frac{P\!\!\!\slash}{M}\Big[\overline{\varphi}
_{_{P_f}}^{++}(\alpha_2P_{fP_\perp}-\alpha_2P_{iP_\perp}+q_{_{P_\perp}})\gamma^\mu\varphi_{_{P_i}}^{++}
(q_{_{P_\perp}})\nonumber\\
&&+\overline{\varphi}
_{_{P_f}}^{++}(\alpha_2P_{fP_\perp}-\alpha_2P_{iP_\perp}+q_{_{P_\perp}})\gamma^\mu\psi_{_{1P_i}}^{-+}
(q_{_{P_\perp}})-\overline{\psi}
_{_{1P_f}}^{-+}(\alpha_2P_{fP_\perp}-\alpha_2P_{iP_\perp}+q_{_{P_\perp}})\gamma^\mu\varphi_{_{P_i}}^{--}
(q_{_{P_\perp}})\nonumber\\
&&-\overline{\psi}
_{_{1P_f}}^{+-}(\alpha_2P_{fP_\perp}-\alpha_2P_{iP_\perp}+q_{_{P_\perp}})\gamma^\mu\varphi_{_{P_i}}^{++}
(q_{_{P_\perp}})+\overline{\varphi}
_{_{P_f}}^{--}(\alpha_2P_{fP_\perp}-\alpha_2P_{iP_\perp}+q_{_{P_\perp}})\gamma^\mu\psi_{_{1P_i}}^{+-}
(q_{_{P_\perp}})\nonumber\\
&&-\overline{\varphi}
_{_{P_f}}^{--}(\alpha_2P_{fP_\perp}-\alpha_2P_{iP_\perp}+q_{_{P_\perp}})\gamma^\mu\varphi_{_{P_i}}^{--}
(q_{_{P_\perp}})\Big]\Big\}.
\end{eqnarray}
For convenience, we re-define
\begin{eqnarray}&&\psi_{_{1P_i}}^{-+}(q_{_{P_\perp}})=\frac{\Lambda^-_1\eta_{_{P_i}}(q_{_{P_\perp}})
\Lambda^+_2}{E+\omega_1+\omega'_1-E'},\\
&&\psi_{_{1P_i}}^{+-}(q_{_{P_\perp}})=\frac{\Lambda^+_1\eta_{_{P_i}}(q_{_{P_\perp}})
\Lambda^-_2}{E-\omega_1-\omega'_1-E'},\\
&&\overline{\psi}_{_{1P_f}}^{-+}(q'_{_{P_\perp}})=\frac{\Lambda'^-_1\overline{\eta}_{_{P_f}}
(q'_{_{P_\perp}})\Lambda'^+_2}{E+\omega_1+\omega'_1-E'},\\
&&\overline{\psi}_{_{1P_f}}^{+-}(q'_{_{P_\perp}})=\frac{\Lambda'^+_1\overline{\eta}_{_{P_f}}
(q'_{_{P_\perp}})\Lambda'^-_2}{E-\omega_1-\omega'_1-E'},
\end{eqnarray}
where $E$ and $E'$ are the energies of the initial and final
diquarks in the center-of-mass frame of the baryon whose momentum
is $P$. Similarly,
\begin{eqnarray}&&M^\mu_2=\int\frac{d^3q_{_{P_\perp}}}{(2\pi)^3}Tr\Big\{-\Big[\overline{\varphi}
_{_{P_f}}^{++}(\alpha_1P_{iP_\perp}-\alpha_1P_{fP_\perp}+q_{_{P_\perp}})\frac{P\!\!\!\slash}{M}\varphi_{_{P_i}}^{++}
(q_{_{P_\perp}})\nonumber\\
&&+\overline{\varphi}
_{_{P_f}}^{++}(\alpha_1P_{iP_\perp}-\alpha_1P_{fP_\perp}+q_{_{P_\perp}})\frac{P\!\!\!\slash}{M}\psi_{_{2P_i}}^{+-}
(q_{_{P_\perp}})-\overline{\psi}
_{_{2P_f}}^{+-}(\alpha_1P_{iP_\perp}-\alpha_1P_{fP_\perp}+q_{_{P_\perp}})\frac{P\!\!\!\slash}{M}\varphi_{_{P_i}}^{--}
(q_{_{P_\perp}})\nonumber\\
&&-\overline{\psi}
_{_{2P_f}}^{-+}(\alpha_1P_{iP_\perp}-\alpha_1P_{fP_\perp}+q_{_{P_\perp}})\frac{P\!\!\!\slash}{M}\varphi_{_{P_i}}^{++}
(q_{_{P_\perp}})+\overline{\varphi}
_{_{P_f}}^{--}(\alpha_1P_{iP_\perp}-\alpha_1P_{fP_\perp}+q_{_{P_\perp}})\frac{P\!\!\!\slash}{M}\psi_{_{2P_i}}^{-+}
(q_{_{P_\perp}})\nonumber\\
&&-\overline{\varphi}
_{_{P_f}}^{--}(\alpha_1P_{iP_\perp}-\alpha_1P_{fP_\perp}+q_{_{P_\perp}})\frac{P\!\!\!\slash}{M}\varphi_{_{P_i}}^{--}
(q_{_{P_\perp}})\Big]\gamma^\mu\Big\}.
\end{eqnarray}
For convenience, we have also re-defined
\begin{eqnarray}&&\psi_{_{2P_i}}^{-+}(q_{_{P_\perp}})=\frac{\Lambda^-_1\eta_{_{P_i}}(q_{_{P_\perp}})
\Lambda^+_2}{E-\omega_1-\omega'_1-E'}\\
&&\psi_{_{2P_i}}^{+-}(q_{_{P_\perp}})=\frac{\Lambda^+_1\eta_{_{P_i}}(q_{_{P_\perp}})
\Lambda^-_2}{E+\omega_1+\omega'_1-E'}\\
&&\overline{\psi}_{_{2P_f}}^{-+}(q'_{_{P_\perp}})=\frac{\Lambda'^-_1\overline{\eta}_{_{P_f}}
(q'_{_{P_\perp}})\Lambda'^+_2}{E-\omega_1-\omega'_1-E'}\\
&&\overline{\psi}_{_{2P_f}}^{+-}(q'_{_{P_\perp}})=\frac{\Lambda'^+_1\overline{\eta}_{_{P_f}}
(q'_{_{P_\perp}})\Lambda'^-_2}{E+\omega_1+\omega'_1-E'},
\end{eqnarray}
and the wavefunctions of the initial and final $0^+$ diquarks are
respectively
\begin{eqnarray}
\varphi_{_{P_i}}(q_{_{P_\perp}})=
\frac{q_{_{P_\perp}}^2(m_1+m_2)(\omega_1-\omega_2)\gamma_0b_4(q_{_{P_\perp}})}{(\omega_1+\omega_2)
(q_{_{P_\perp}}^2-m_1m_2-\omega_1\omega_2)}\nonumber\\+\frac{q_{_{P_\perp}}^2(m_1+m_2)b_3(q_{_{P_\perp}})}
{q_{_{P_\perp}}^2-m_1m_2-\omega_1\omega_2}
+q\!\!\!\slash_{_{P_\perp}}b_3(q_{_{P_\perp}})+\gamma_0
q\!\!\!\slash_{_{P_\perp}}b_4(q_{_{P_\perp}}),
\end{eqnarray}
\begin{eqnarray}\overline{\varphi}_{_{P_f}}(q'_{_{P_\perp}})=
-\frac{q_{_{P_\perp}}'^2(m_1+m_2)(\omega'_1-\omega'_2)\gamma_0b_4(q'_{_{P_\perp}})}{(\omega'_1+\omega'_2)
(q_{_{P_\perp}}'^2-m_1m_2-\omega'_1\omega'_2)}\nonumber\\-\frac{q_{_{P_\perp}}^2(m_1+m_2)b_3(q'_{_{P_\perp}})}
{q_{_{P_\perp}}'^2-m_1m_2-\omega'_1\omega'_2}
-q\!\!\!\slash'_{_{P_\perp}}b_3(q'_{_{P_\perp}})+\gamma_0
q\!\!\!\slash'_{_{P_\perp}}b_4(q'_{_{P_\perp}}).
\end{eqnarray}

2. The form factors at the effective vertex for vector diquarks
coupling to gluons.

This effective coupling has been discussed by some authors
\cite{anselmino1,guo} and the effective vertex has the following
form
\begin{equation}\Gamma^{\alpha\mu\beta,a}_{A g A} =-ig_s\frac{\lambda^a}{2}\Big[G_1(Q^2)(P_f+P_i)^\mu
g^{\alpha\beta}-G_2(Q^2)(P_f^\alpha g^{\mu\beta}+P_i^\beta
g^{\mu\alpha})+G_3(Q^2)(P_f+P_i)^\mu P^\alpha_fP^\beta_i\Big],
\end{equation}
which in the BSE approach can be further expressed as
\begin{equation}
\Gamma^{\alpha\mu\beta,a}_{A g
A}=-ig_s\frac{\lambda^a}{2}(M^{\alpha\mu\beta}_1+M^{\alpha\mu\beta}_2),
\end{equation}
where
\begin{eqnarray}
M^{\alpha\mu\beta}_1=-\int\frac{d^4q}{(2\pi)^4}Tr\Big[\overline{\chi}^\beta_{_{P_f}}(q')
\gamma^\mu\chi^\alpha_{_{P_i}}(q) S^{-1}_F(p_2)\Big],\label{eqm11}\\
M^{\alpha\mu\beta}_2=-\int\frac{d^4q}{(2\pi)^4}Tr\Big[\overline{\chi}^\beta_{_{P_f}}(q')
S^{-1}_F(p_1)\chi^\alpha_{_{P_i}}(q)\gamma^\mu \Big],\label{eqm12}
\end{eqnarray}
and
\begin{eqnarray}\chi^\alpha_{_{P_i}}(q)\epsilon^\lambda_\alpha=\chi_{_{P_i}}(q),
\end{eqnarray}
where $\epsilon^\lambda$ is the polarization vector of the vector
diquark. In analog to the method given in last section, we can
simplify $M^{\alpha\mu\beta}$ as follows:
\begin{eqnarray}&&M^{\alpha\mu\beta}(\Gamma_{qgq})=\int\frac{d^3q_{_{P_\perp}}}{(2\pi)^3}Tr\Big\{-\frac{P\!\!\!\slash}{M}\Big[\overline{
\varphi}_{_{P_f}}^{\beta++}(\alpha_2P_{fP_\perp}-\alpha_2P_{iP_\perp}+q_{_{P_\perp}})\gamma^\mu\varphi_{_{P_i}}^{\alpha++}
(q_{_{P_\perp}})\nonumber\\
&&+\overline{\varphi}
_{_{P_f}}^{\beta++}(\alpha_2P_{fP_\perp}-\alpha_2P_{iP_\perp}+q_{_{P_\perp}})\gamma^\mu\psi_{_{1P_i}}^{\alpha-+}
(q_{_{P_\perp}})-\overline{\psi}
_{_{1P_f}}^{\beta-+}(\alpha_2P_{fP_\perp}-\alpha_2P_{iP_\perp}+q_{_{P_\perp}})\gamma^\mu\varphi_{_{P_i}}^{\alpha--}
(q_{_{P_\perp}})\nonumber\\
&&-\overline{\psi}
_{_{1P_f}}^{\beta+-}(\alpha_2P_{fP_\perp}-\alpha_2P_{iP_\perp}+q_{_{P_\perp}})\gamma^\mu\varphi_{_{P_i}}^{\alpha++}
(q_{_{P_\perp}})+\overline{\varphi}
_{_{P_f}}^{\beta--}(\alpha_2P_{fP_\perp}-\alpha_2P_{iP_\perp}+q_{_{P_\perp}})\gamma^\mu\psi_{_{1P_i}}^{\alpha+-}
(q_{_{P_\perp}})\nonumber\\
&&-\overline{\varphi}
_{_{P_f}}^{\beta--}(\alpha_2P_{fP_\perp}-\alpha_2P_{iP_\perp}+q_{_{P_\perp}})\gamma^\mu\varphi_{_{P_i}}^{\alpha--}
(q_{_{P_\perp}})\Big]-\Big[\overline{\varphi}
_{_{P_f}}^{\beta++}(\alpha_1P_{iP_\perp}-\alpha_1P_{fP_\perp}+q_{_{P_\perp}})\frac{P\!\!\!\slash}{M}\varphi_{_{P_i}}^{\alpha++}
(q_{_{P_\perp}})\nonumber\\
&&+\overline{\varphi}
_{_{P_f}}^{\beta++}(\alpha_1P_{iP_\perp}-\alpha_1P_{fP_\perp}+q_{_{P_\perp}})\frac{P\!\!\!\slash}{M}\psi_{_{2P_i}}^{\alpha+-}
(q_{_{P_\perp}})-\overline{\psi}
_{_{2P_f}}^{\beta+-}(\alpha_1P_{iP_\perp}-\alpha_1P_{fP_\perp}+q_{_{P_\perp}})\frac{P\!\!\!\slash}{M}\varphi_{_{P_i}}^{\alpha--}
(q_{_{P_\perp}})\nonumber\\
&&-\overline{\psi}
_{_{2P_f}}^{\beta-+}(\alpha_1P_{iP_\perp}-\alpha_1P_{fP_\perp}+q_{_{P_\perp}})\frac{P\!\!\!\slash}{M}\varphi_{_{P_i}}^{\alpha++}
(q_{_{P_\perp}})+\overline{\varphi}
_{_{P_f}}^{\beta--}(\alpha_1P_{iP_\perp}-\alpha_1P_{fP_\perp}+q_{_{P_\perp}})\frac{P\!\!\!\slash}{M}\psi_{_{2P_i}}^{\alpha-+}
(q_{_{P_\perp}})\nonumber\\
&&-\overline{\varphi}
_{_{P_f}}^{\beta--}(\alpha_1P_{iP_\perp}-\alpha_1P_{fP_\perp}+q_{_{P_\perp}})\frac{P\!\!\!\slash}{M}\varphi_{_{P_i}}^{\alpha--}
(q_{_{P_\perp}})\Big]\gamma^\mu\Big\},
\end{eqnarray}
where
\begin{eqnarray}&&\varphi_{_{P_i}}(q_{_{P_\perp}})=\epsilon^\lambda_{\perp\mu}\Big\{q^\mu_{_{P_\perp}}
\Big[f_1(q_{_{P_\perp}})+\gamma_0\frac{\omega_2-\omega_1}
{\omega_1+\omega_2}f_8(q_{_{P_\perp}})+\gamma_0\frac{q_{_{P_\perp}}^2-m_1m_2-\omega_1
\omega_2}{M(m_1+m_2)}f_4(q_{_{P_\perp}})\nonumber\\
&&+\frac{q\!\!\!\slash_{_{P_\perp}}[M\omega_1f_5(q_{_{P_\perp}})+(m_2\omega_1-m_1\omega_2)f_1(q_{_{P_\perp}})]}
{q_{_{P_\perp}}^2(\omega_1+\omega_2)}+\frac{\gamma_0q\!\!\!\slash_{_{P_\perp}}
f_4(q_{_{P_\perp}})}{M}\Big]+M\gamma^\mu
f_5(q_{_{P_\perp}})\nonumber\\
&&+M\gamma^\mu\gamma_0
\frac{m_1\omega_2-m_2\omega_1}{M(w_1+w_2)}f_8(q_{_{P_\perp}})
-(q\!\!\!\slash_{_{P_\perp}}\gamma^\mu-q^\mu_{_{P_\perp}})\frac{M(m_1\omega_2
+m_2\omega_1)}{q_{_{P_\perp}}^2(\omega_1+\omega_2)}f_5(q_{_{P_\perp}})\nonumber\\
&&+\gamma_0(q^\mu_{_{P_\perp}}-\gamma^\mu
q\!\!\!\slash_{_{P_\perp}})f_8(q_{_{P_\perp}})]\Big\}\gamma_5=\varphi^\mu_{_{P_i}}(q_{_{P_\perp}})
\epsilon^\lambda_{\perp\mu},
\end{eqnarray}
and
\begin{eqnarray}&&\overline{\varphi}_{_{P_f}}(q'_{_{P_\perp}})=\epsilon^\lambda_{\perp\mu}\Big\{q'^\mu_{_{P_\perp}}
\Big[f_1(q'_{_{P_\perp}})-\gamma_0\frac{\omega_2-\omega_1}
{\omega'_1+\omega'_2}f_8(q'_{_{P_\perp}})-\gamma_0\frac{q'^2_{_{P_\perp}}-m_1m_2-\omega'_1
\omega'_2}{M(m_1+m_2)}f_4(q'_{_{P_\perp}})\nonumber\\
&&-\frac{q\!\!\!\slash'_{_{P_\perp}}[M\omega'_1f_5(q_{_{P_\perp}})+(m_2\omega'_1-m_1\omega'_2)f_1(q'_{_{P_\perp}})]}
{q'^2_{_{P_\perp}}(\omega'_1+\omega'_2)}+\frac{q\!\!\!\slash'_{_{P_\perp}}\gamma_0
f_4(q'_{_{P_\perp}})}{M}\Big]-M\gamma^\mu
f_5(q_{_{P_\perp}})\nonumber\\
&&+M\gamma_0\gamma^\mu
\frac{m_1\omega_2-m_2\omega_1}{M(w_1+w_2)}f_8(q_{_{P_\perp}})
+(q^\mu_{_{P_\perp}}-\gamma^\mu
q\!\!\!\slash_{_{P_\perp}})\frac{M(m_1\omega_2
+m_2\omega_1)}{q_{_{P_\perp}}^2(\omega_1+\omega_2)}f_5(q_{_{P_\perp}})\nonumber\\
&&+\gamma_0(q^\mu_{_{P_\perp}}-q\!\!\!\slash_{_{P_\perp}}\gamma^\mu
)f_8(q_{_{P_\perp}})]\Big\}\gamma_5=\overline{\varphi}^\mu_{_{P_i}}(q_{_{P_\perp}})
\epsilon^\lambda_{\perp\mu}.
\end{eqnarray}

\subsection{The form factors at the effective vertices of diquark coupling to $\gamma$, $Z^0$, $W^{\pm}$}

 \hspace{0.5cm}   1. The effective current for a scalar diquark coupling to a photon
can be written as,
\begin{eqnarray}\Gamma^\mu_{s\gamma s}&&=-ieG(Q^2)(P_i+P_f)^\mu\nonumber\\
&&=-ie(e_1M_1^\mu+e_2M_2^\mu),
\end{eqnarray}
where $e_1$ and $e_2$ are the charges of the quarks  in the
diquark with momenta $p_1$ and $p_2$ respectively.

2. The effective current for a vector diquark coupling to a photon
is

\begin{eqnarray}\Gamma^{\alpha\mu\beta}_{A\gamma A}&&=-ie\Big[G_1(Q^2)(P_f+P_i)^\mu
g^{\alpha\beta}\nonumber\\&&-G_2(Q^2)(P_f^\alpha
g^{\mu\beta}+P_i^\beta g^{\mu\alpha})+G_3(Q^2)(P_f+P_i)^\mu
P^\alpha_fP^\beta_i\Big]\nonumber\\
&&=-ie(e_1M^{\alpha\mu\beta}_1+e_2M^{\alpha\mu\beta}_2).
\end{eqnarray}

3. The effective current for a scalar diquark coupling to $Z^0$
can be written as
\begin{eqnarray}
\Gamma^\mu_{sZs}&=&-igG(Q^2)(P_f+P_i)^\mu\nonumber\\
&=&-ig(M^\mu_1+M^\mu_2).
\end{eqnarray}

4. The effective current for a vector diquark coupling to $Z^0$ is

\begin{eqnarray}\Gamma^{\alpha\mu\beta}_{AZA}&&=-ig\Big\{G_1(Q^2)(P_f+P_i)^\mu
g^{\alpha\beta}-G_2(Q^2)(P_f^\alpha g^{\mu\beta}+P_i^\beta
g^{\mu\alpha})\nonumber\\&&+G_3(Q^2)(P_f+P_i)^\mu
P^\alpha_fP^\beta_i\nonumber\\&&-iG_{4}(Q^2)\epsilon^{\alpha\mu\beta\sigma}
(P_f+P_i)_\sigma-iG_{5}(Q^2)\epsilon^{\alpha\mu\beta\sigma}
(P_f-P_i)_\sigma\nonumber\\&&-iG_{6}(Q^2)\epsilon^{\alpha\beta\sigma\rho}P_{f\sigma}P_{i\rho}(P_f+P_i)^\mu
-iG_{7}(Q^2)\epsilon^{\alpha\beta\sigma\rho}P_{f\sigma}P_{i\rho}(P_f-P_i)^\mu\nonumber\\
&&-iG_{8}(Q^2)(\epsilon^{\mu\beta\sigma\rho}P_{f\sigma}P_{i\rho}P_f^\alpha+\epsilon^{\alpha\mu\sigma\rho}
P_{f\sigma}P_{i\rho}P_i^\beta)\nonumber\\&&-iG_{9}(Q^2)(\epsilon^{\mu\beta\sigma\rho}P_{f\sigma}P_{i\rho}P_f^\alpha
-\epsilon^{\alpha\mu\sigma\rho}P_{f\sigma}P_{i\rho}P_i^\beta)\Big\}\nonumber\\
=&&-ig(M_1^{\alpha\mu\beta}+M_2^{\alpha\mu\beta}).
\end{eqnarray}

5. The effective current for a scalar diquark coupling to $W^\pm$
is
\begin{eqnarray}
\Gamma^{\mu}_{sWs}&&=-ig\Big[G_1(Q^2)(P_f+P_i)^\mu+G_2(Q^2)(P_f-P_i)^\mu\Big]\nonumber\\
&&=-ig(M_1^\mu+M_2^\mu).
\end{eqnarray}

6. The effective current for a vector-diquark coupling to $W^\pm$
is
\begin{eqnarray}\Gamma^{\alpha\mu\beta}_{AWA}&&=-ig\Big\{G_1(Q^2)(P_f+P_i)^\mu
g^{\alpha\beta}-G_2(Q^2)(P_f^\alpha g^{\mu\beta}+P_i^\beta
g^{\mu\alpha})\nonumber\\&&-G_3(Q^2)(P_f^\alpha
g^{\mu\beta}-P_i^\beta g^{\mu\alpha})+G_4(Q^2)(P_f+P_i)^\mu
P^\alpha_fP^\beta_i\nonumber\\&&+G_5(Q^2)(P_f-P_i)^\mu
g^{\alpha\beta}+G_6(Q^2)(P_f-P_i)^\mu
P^\alpha_fP^\beta_i\nonumber\\&&-iG_{7}(Q^2)\epsilon^{\alpha\mu\beta\sigma}
(P_f+P_i)_\sigma-iG_{8}(Q^2)\epsilon^{\alpha\mu\beta\sigma}
(P_f-P_i)_\sigma\nonumber\\&&-iG_{9}(Q^2)\epsilon^{\alpha\beta\sigma\rho}P_{f\sigma}P_{i\rho}(P_f+P_i)^\mu
-iG_{10}(Q^2)\epsilon^{\alpha\beta\sigma\rho}P_{f\sigma}P_{i\rho}(P_f-P_i)^\mu\nonumber\\
&&-iG_{11}(Q^2)(\epsilon^{\mu\beta\sigma\rho}P_{f\sigma}P_{i\rho}P_f^\alpha+\epsilon^{\alpha\mu\sigma\rho}
P_{f\sigma}P_{i\rho}P_i^\beta)\nonumber\\&&-iG_{12}(Q^2)(\epsilon^{\mu\beta\sigma\rho}P_{f\sigma}P_{i\rho}P_f^\alpha
-\epsilon^{\alpha\mu\sigma\rho}P_{f\sigma}P_{i\rho}P_i^\beta)\Big\}\nonumber\\
&&=-ig(M_1^{\alpha\mu\beta}+M_2^{\alpha\mu\beta}).
\end{eqnarray}

7. The effective current for a vector diquark-$W^\pm$-scalar
diquark coupling is written as
\begin{eqnarray}\Gamma^{\alpha\mu}_{AWs}&&=-ig\Big[G_1(Q^2)P^\alpha_fP^\mu_i+G_2(Q^2)g^{\alpha\mu}
-iG_3(Q^2)\epsilon^{\alpha\mu\sigma\rho}P_{f\sigma}P_{i\rho}
\Big]\nonumber\\
&&=-ig(M_1^{\alpha\mu}+M_2^{\alpha\mu}),
\end{eqnarray}where
\begin{eqnarray}M_1^{\alpha\mu}=-\int\frac{d^4q}{(2\pi)^4}Tr\Big[\overline{\chi}_{_{P_f}}(q')
V^\mu_{q_1Wq'_1}\chi^\alpha_{_{P_i}}(q) S^{-1}_F(p_2)\Big]\\
M_2^{\alpha\mu}=-\int\frac{d^4q}{(2\pi)^4}Tr\Big[\overline{\chi}_{_{P_f}}(q')
S^{-1}_F(p_1)\chi^\alpha_{_{P_i}}(q)V^\mu_{q_2Wq'_2}\Big].
\end{eqnarray}

\subsection{The form factors at the effective vertices of diquark coupling to $\pi$ mesons}

To complete the picture, we also discuss the effective couplings
of scalar or vector diquarks and $\pi$ mesons in terms of the
chiral Lagrangian, which has been widely adopted in the studies of
the effective quark-meson-quark couplings.

The effective coupling of quark and $\pi-$meson is of the form
\cite{jin}
\begin{eqnarray}\Gamma_{q\pi
q}=\frac{g_q}{f_\pi}\gamma_5k\!\!\!\slash,\end{eqnarray} where $k$
is the momentum of the pion. We obtain the effective coupling
vertex of $1^+$ diquark and $\pi-$meson as following
\begin{eqnarray}
\Gamma_{A\pi
A}=\epsilon^1_{\alpha}\epsilon^{2*}_{\beta}\Gamma^{\alpha\beta}_{A\pi
A}
\end{eqnarray}
where $\epsilon^{1,2}$ are the polarization vectors of the two
axial-vector diquarks and
\begin{eqnarray} \Gamma^{\alpha\beta}_{A\pi
A}&&=i\frac{g_q}{f_\pi}F_{A\pi A}(Q^2)\epsilon^{\alpha\beta\sigma\rho}P_{f\sigma}P_{i\rho}\nonumber\\
&&=i\frac{g_q}{f_\pi}[M_1^{\alpha\beta}+M_2^{\alpha\beta}]
\end{eqnarray}where
\begin{eqnarray}M_1^{\alpha\beta}=-\int\frac{d^4q}{(2\pi)^4}Tr\Big[\overline{\chi}^\beta_{_{P_f}}(q')
\Gamma_{q\pi
q}\chi^\alpha_{_{P_i}}(q) S^{-1}_F(p_2)\Big]\\
M_2^{\alpha\beta}=-\int\frac{d^4q}{(2\pi)^4}Tr\Big[\overline{\chi}^\beta_{_{P_f}}(q')
S^{-1}_F(p_1)\chi^\alpha_{_{P_i}}(q)\Gamma_{q\pi q}\Big],
\end{eqnarray}

\section{numerical results}

The formulas derived above are for form factors at the effective
vertices of any diquark which couples to gauge bosons. However, as
indicated in the introduction, the diquark picture only works
without any doubt for the heavy diquarks. For light diquark, or
heavy-light diquark, the relativistic effects may be crucial,
therefore in this work, to avoid any ambiguity, we only
numerically evaluate the form factors of the $bc-$diquark coupling
to gauge bosons. There are both scalar and axial vector $bc$
diquark (here we do not concern the orbital excited states),
whereas there is only axial vector diquarks for $bb$ and $cc$.

For applying the form factors under consideration to transition
processes, where baryons are involved, we need to present the
numerical values which are computed in terms of the programs
developed by Chang et al. The input parameters are
\cite{yu,wang1}: $ m_c=1.7553\; {\rm GeV},\;\; m_b=5.224\; {\rm
GeV},\;\; \lambda=0.20\; {\rm GeV}^2 , \Lambda_{QCD}=0.26\; {\rm
GeV},\;\; a=2.71828\;\;,{\rm}\;\; \alpha=0.06\; {\rm GeV},\;\;
\beta=0.5,\;\;V_0=-0.3\;\; {\rm GeV}$.

We plot the dependence of the form factors for scalar diquark
coupling to gluon, photon and $Z^0$ on $Q^2$  in Figs. 2, 3 and 4
respectively.
\begin{figure}[tbp]
\begin{center}
\scalebox{1.1}{\includegraphics{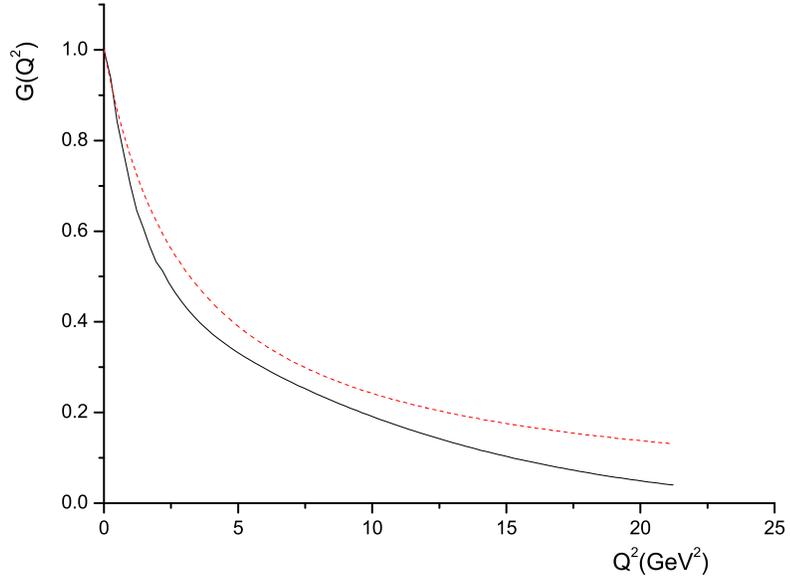}}\end{center}
    \caption{The form factor for the effective vertex of scalar diquark coupling to gluons. The solid line
    is the result calculated in terms of the BSE and the dashed-line corresponds to the form factor which is phenomenologically
    introduced by the authors of Ref. \cite{anselmino1}}
    \label{pic2}
\end{figure}
\begin{figure}[tbp]
\begin{center}
\scalebox{1.1}{\includegraphics{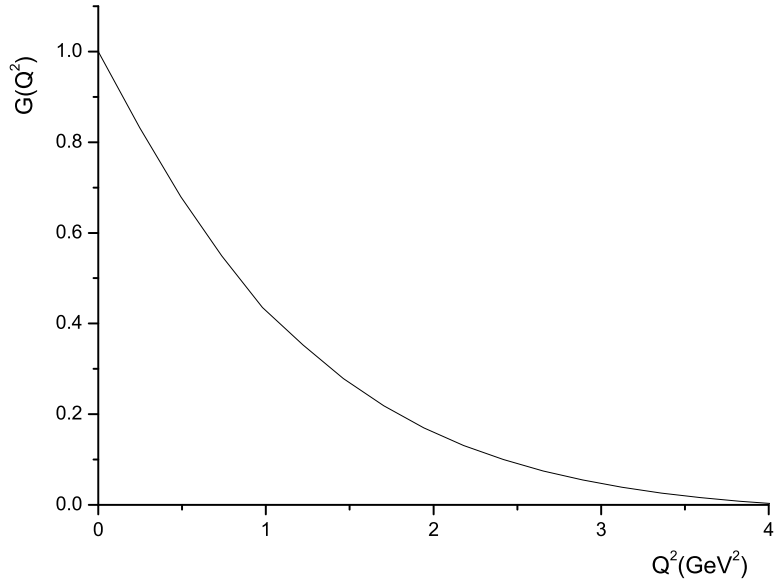}}
    \end{center}
    \caption{The form factor for the effective vertex of scalar diquark coupling to photon}
    \label{pic4}
\end{figure}
\begin{figure}[tbp]
\begin{center}
\scalebox{1.1}{\includegraphics{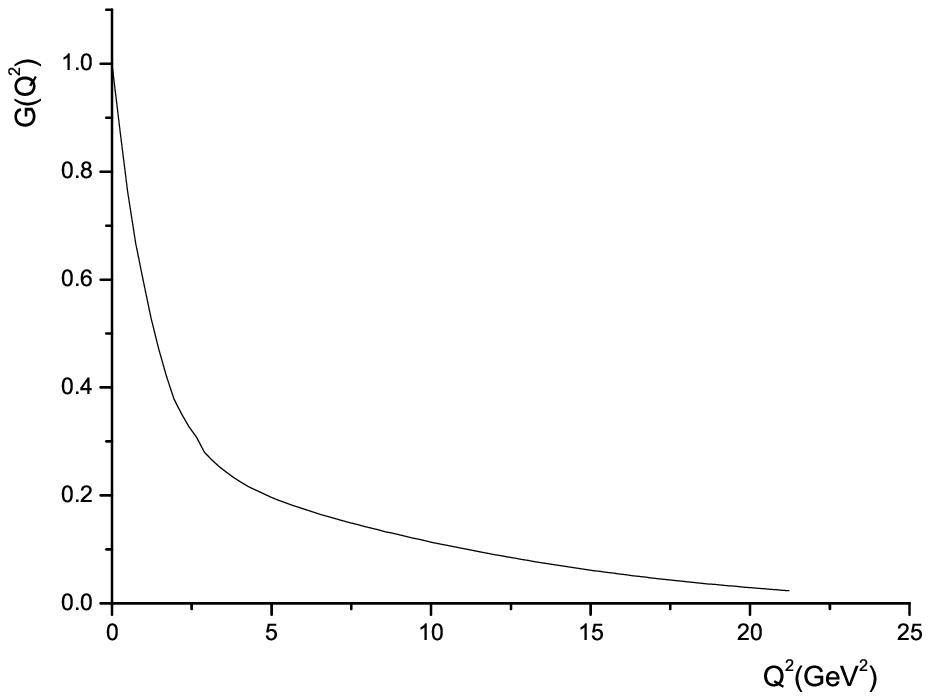}}
    \end{center}
    \caption{The form factor for the effective vertex of  scalar diquark coupling to
    $Z^0$.}
    \label{pic5}
\end{figure}
It is noted that the three independent form factors for the
effective vertex of a vector diquark coupling to a gluon are
simply attributed into only one form factor by reasonable physical
considerations in \cite{Korner}. For a comparison, to extract the
form factor, in our calculations, we employ the relations given in
the reference and adopt the corresponding spin-functions of
baryons, then we plot the form factors  at the effective vertex of
vector diquark coupling to gluon in Fig. 5.
\begin{figure}[tbp]
\begin{center}
\scalebox{1.1}{\includegraphics{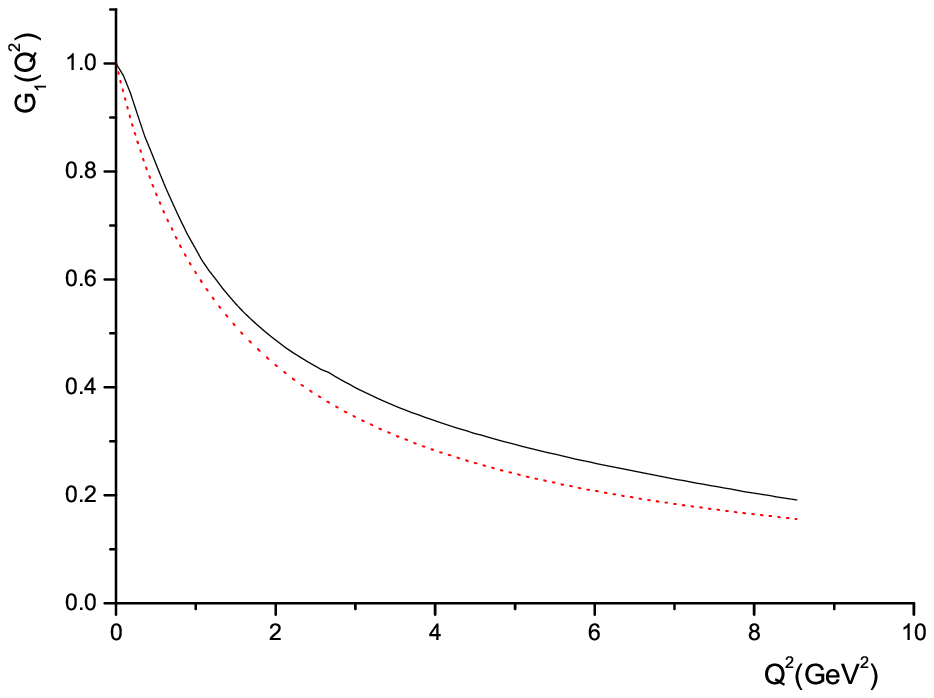}}
    \end{center}
    \caption{The form factors for the effective vertex of vector diquark coupling to gluon. The solid line
    corresponds to the result obtained by the BSE method and the dashed line corresponds to
    the phenomenological form factor given in Ref. \cite{anselmino1}}
    \label{pic6}
\end{figure}

Moreover, since W-emission is accompanied by a flavor change, the
initial diquark $bc$ should transit into $cc+W^-$ or $bs+W^+$. The
situation is slightly more complicated. Therefore, for only
illustration of the behavior of the diquark form factors, we do
not numerically evaluate $\Gamma_{DWD'}$ where $D$ and $D'$ have
different flavors and in our following work, we will present them
along with other form factors such as $\Gamma_{AGS}$ etc.

\section{Summary and discussions}

In this work we derive the form factors for the effective vertices
of scalar and vector diquarks coupling to gauge bosons, $g$,
$\gamma$, $W^{\pm}$ and $Z^0$,  as well as to the $\pi-$meson. We
carry out our derivations in the framework of the Bethe-Salpeter
equation. Even though the BSE is established on the quantum field
theory and its validity is not dubious, there still exist some
uncertainties when it is applied to deal with practical problems.
First the kernel in the equation is not derived based on the
fundamental principles, namely because the non-perturbative QCD
effects are taken into account, corresponding interaction must be
phenomenologically introduced. In this work, we adopt the simple
Cornell potential as the kernel. Then for solving the equation,
one needs to adopt the instantaneous approximation, and then the
original Lorentz invariance is lost. Therefore, for the systems
where the relativistic corrections are important, the
approximation would bring up large errors. However, for the system
especially the systems where only heavy quarks are involved, the
results are more reliable. In our earlier works about the spectra
of diquarks, it was indicated that if there are light quark
constituents, one may make certain modifications. One efficient
way is to consider the BSE and the Dyson-Schwinger equation
simultaneously\cite{dse}. This approach might alleviate the
severity of the error, but cannot finally eliminate all
shortcomings in the framework. In our later work, we are going to
deal with the light diquarks and then estimate the errors. So far,
even though we know the origin of the uncertaities, we cannot
quantitatively estimate their magnitudes.

On the other side, even though the framework has some problems, it
is applicable to the processes where baryons, especially heavy
baryons are concerned. Since the diquark picture greatly
simplifies the whole calculation and also has achieved remarkable
success in phenomenology, one has reason to believe that the
diquark picture is suitable for dealing with the baryon production
or decay processes. Diquark is definitely not a point-like
particle, therefore a form factor(s) at the effective vertex of
diquark coupling to gauge bosons and even pions can partly
compensate the effect of the inner structure of diquarks. We
employ the BSE to derive the form factors. Fortunately, the recent
high energy experiments provide more and more accurate information
about the baryon structure, and we can wait for more data to test
our derivation and find the applicability of this approach.

For a demonstration, we would like to compare the asymptotic
behavior of the form factor $F(Q^2)$, which is derived in this
work with its phenomenological form given by the authors of
Ref.\cite{anselmino1} in Fig. \ref{pic2}. In Fig. 2, we compare
the result obtained in terms of the BSE which is represented by
the solid line, and the dashed line corresponds to the
phenomenological form factor $F(Q^2)=\frac{Q^2_0}{Q^2_0+Q^2}$. It
is noted that the form factor introduced in Ref.\cite{anselmino1}
is for an $ud-$scalar diquark. Since the form factor is introduced
phenomenologically by fitting data, the relativistic effects are
included in the parameters. QCD is flavor blind, so that we
believe that the form of the form factor for $bc$ and $ud$ must be
similar except that the parameter $Q_0^2$ which is related to the
constituents of the diquark may be different, at least their
tendency behavior must be similar. Therefore this comparison is
qualitatively significant, but small deviations would be expected.

$F(Q^2)$ decreases monotonically as $Q^2$ becomes large and
approaches to zero rather quickly. The authors of
Ref.\cite{anselmino1} introduced a phenomenological form factor as
$F(Q^2)=\frac{Q^2_0}{Q^2_0+Q^2}$, where $Q_0$ is a parameter
determined  as   $Q_0^2\sim 3.2$ GeV$^2$ by fitting data
\cite{anselmino1}. The form is obviously understandable. The form
factors should be normalized to unity as $Q^2\to 0$, i.e. as one
looks at the diquark from a far distance, the form factor becomes
a unity, whereas as $Q^2\to\infty$, the inspector then penetrates
into the diquark, so that he would see the individual quarks
instead of the whole and the diquark picture no longer holds and
mathematically it is required to approach zero as $Q^2\to \infty$.
The form factor obtained in terms of the BSE generally coincides
with the picture.

Since diquark is a boson of color-anti-triplet, it cannot exist as
a physical object, but a constituent in baryon, just like a quark
in meson. Besides, it resides in a bound state, therefore must be
off-shell, but for a not-very-tight bound state, it can be treated
as a physical object which is approximately on its mass shell.
Thus one can use the wavefunctions of the diquark for calculating
the form factors, but obviously certain errors may be caused. All
the form factors obtained in this work cannot be directly tested
because diquark does not exist as an individual. To test their
validity, one needs to apply them into the practical processes
where baryons are concerned. Therefore, in our next work, we will
calculate the production and decay rates of the processes where
baryons are involved, in terms of the form factors derived here
and let data confirm or negate this picture, if conclusion is
positive, the accuracy degree will also be determined by the data.

As a conclusion, the diquark picture is reasonable and can be
applied to study the processes where baryons are involved,
especially for the baryons with two heavy quarks, as long as
suitable form factors are included. The form factors derived in
terms of the BSE are consistent with that obtained by fitting
data, namely, they are applicable in practical calculations.
However, for the diquark including two light quarks or that
including one light and one heavy quarks, the errors in the
calculations may be large. For achieving form factors for diquarks
which are composed of only light quarks or one light and one heavy
quarks should be studied in a more complicated framework
which would be the goal of our next work.\\

\noindent{Acknowledgement:}

We thank C.-H. Chang for helpful discussions. This work is partly
supported by the National Natural Science Foundation of China
(NNSFC). \\

\end{document}